 \documentclass[final,5p,times,twocolumn,sort&compress]{elsarticle}

 \usepackage{graphicx,amssymb,amsmath,amsthm,lineno}
 \usepackage{dcolumn,multirow}
 \usepackage[center]{subfigure}
 \usepackage{threeparttable}
 \usepackage[mathscr]{euscript}
 \usepackage{mathtools, cuted}
 \usepackage[english]{babel}
 \usepackage{lipsum}

\biboptions{square}

\begin{document}

\begin{frontmatter}



\title{Random packing dynamics of  $\Sigma_{2v}(2\pi/3)$-triplets}


\author[ufsa]{Carlos Handrey Araujo Ferraz\corref{cor1}}
\ead{handrey@ufersa.edu.br}
\cortext[cor1]{Corresponding author}
\address[ufsa]{Exact and Natural Sciences Center, Universidade Federal Rural do Semi-\'Arido-UFERSA, PO Box 0137, CEP 59625-900, Mossor\'o, RN, Brazil}

\begin{abstract}
 In this letter, we used a combination of DEM and the multi-sphere method to investigate the random packing dynamics of $\Sigma_{2v}(2\pi/3)$-triplets. These triplets consist of three overlapping primary spheres, forming a bent structure with a bond angle of $2\pi/3$ and belonging to the $C_{2v}$ symmetry group. The selection of this specific structure was motivated by its similarity to molecules such as water, which displays crucial physicochemical properties and finds extensive application in various fields. To ensure non-overlapping particles at the beginning of the simulations, the rectangular confinement box was divided into basic cells. Each triplet was then inserted into a basic cell with a random orientation. After that, the system is allowed to settle under gravity towards the bottom of the box. An implicit leapfrog algorithm with quaternion acceleration was used to numerically integrate the rotational motion equations. By assuming a molecular approach, we account for the long-range cohesive forces using a Lennard-Jones (LJ)-like potential. The packing processes are studied assuming different long-range interaction strengths. We performed statistical calculations of the different quantities studied including packing density, radial distribution function, and orientation pair correlation function. In addition, the force probability distributions in the random packing structures have been analyzed.
\end{abstract}


\end{frontmatter}


\section{Introduction \label{sec:int}}

The importance of particle packing research in materials engineering and science is undeniable. It has implications for various fields such as polymer and drug development, food science, amorphous materials, glasses, ceramic compounds, and sintering processes, among others. The physico-mechanical properties of packed structures are critical for understanding their response to axial compressions, stress-strain response, and predicting potential failures or gaps inside the aggregates.

With the advent of fast computers, the discrete element method (DEM)~\cite{Cundall1979} has been utilized to study the behavior of granular matter by calculating forces and torques between particles. This allows for obtaining trajectories, spins, and orientations of all interacting particles, thereby enabling the assessment of the entire system's temporal evolution. While the spherical shape has traditionally been the preferred choice in granular matter research due to its ease in detecting interparticle contacts and calculating forces through particle deformation during compression, real-world processes often involve particles with non-spherical shapes. Examples include polymeric powder, solid fuel, and manufactured catalyst carriers, which exhibit high irregularity. The shape of these components can significantly impact various properties such as packing properties, powder flowability, interaction with fluids, pigment coverage, and more.
 
Previous numerical studies~\cite{Ferellec2010,Wachs2012,Williams2014,He2017,Alireza2018} have demonstrated that particle shape strongly influences the mechanical behavior of particles, and these findings are supported by experiments~\cite{Zou1996, Villarruel2000, Donev2004}. Non-spherical particles have shown a reduced tendency to rotate compared to spherical ones, resulting in higher shear strength under axial compression~\cite{Zhao2015, Alireza2018}. This enhanced shear strength is attributed to interlocking between particles, which provides rolling resistance. Furthermore, experimental and computational studies have shown that the packing behavior of non-spherical particles differs significantly from that of spherical particles. In general, non-spherical particles give a lower random packing density and coordination number than spherical particles due to higher interparticle friction. Moreover, non-spherical particles exhibit distinct patterns of contact points, often characterized by distributed regions of contact rather than singular points. 

Over the years, there has been increasing interest in the dynamics of random packing involving non-spherical particles~\cite{Lu2015}. However, a complete understanding of how the size, shape, nature of particles, and the packing method used influence both the dynamics and the formed structures remains an open question. Most investigations have primarily focused on spheroids/ellipsoids~\cite{Zhou2011,Baram2012,Zheng2013,He2017,Li2018}, cylinders~\cite{Kodam2010,Guo2012}, and polyhedra~\cite{Boon2012, Nassauer2013}. However, there have been limited reports on studies involving particles with a more molecular aspect, such as diatomic or triatomic molecules. Therefore, it is crucial to comprehend how these fundamental particles interact in random packing processes. Additionally, it is important to investigate the impact of long-range forces, including electrostatic and van der Waals forces, on both the dynamics and inherent properties of the resulting packings. These forces are typically encountered in processes involving such particles. A recent study on the packing of binary particles~\cite{Ferraz2021} has revealed the presence of shield effects against long-range forces. We wonder whether similar effects can also be observed in binary particle packings with some structural constraints. 

\begin{figure}
	\centering
	\includegraphics[scale=0.36, angle=0]{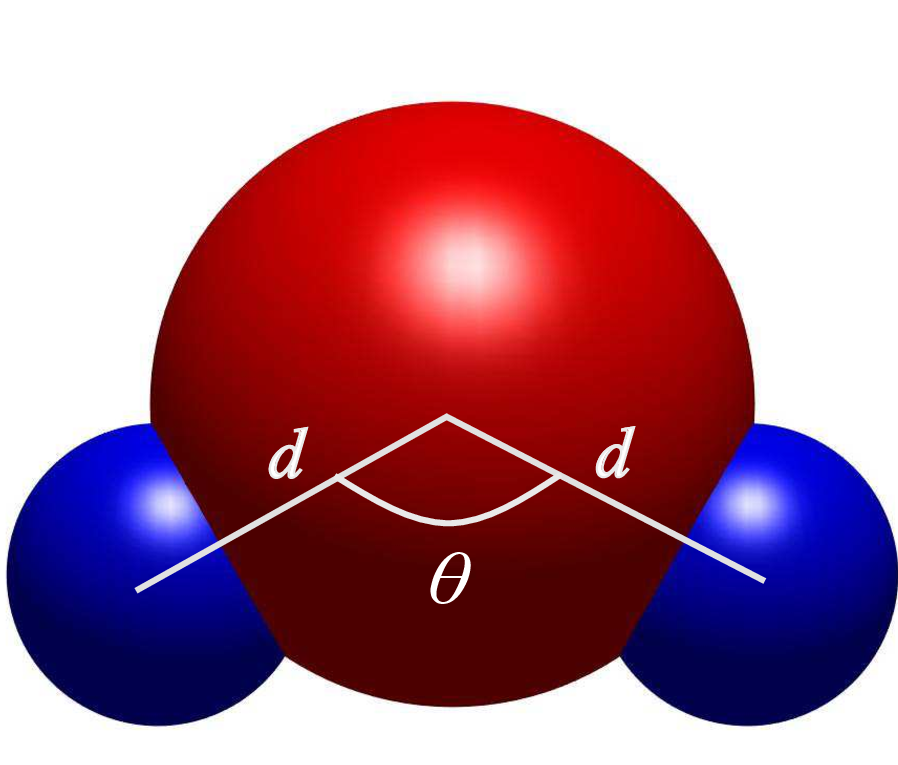}
\caption{Representation of the $\Sigma_{2v}(2\pi/3)$-triplet obtained from the overlapping of two primary spheres of radius $a=0.10 \, \mu m$ (shown in blue) with one sphere of radius $b=0.20 \, \mu m$ (shown in red), arranged in a V-shaped formation. Above $d=3/4\,(a + b)$ and $\theta=2\pi/3$.}\label{fig:01}
\end{figure}

In the present study, we use a combination of Discrete Element Method (DEM) and the multi-sphere method~\cite{Favier1999,Thornton1999} to investigate the random packing dynamics of  $\Sigma_{2v}(2\pi/3)$-triplets. The triplets are comprised of three overlapping primary spheres, resulting in a bent structure with a bond angle of $2\pi/3$ and belonging to the $C_{2v}$ symmetry group (see Fig.~\ref{fig:01}). We chose this structural arrangement based on molecules such as water, which exhibits critical physicochemical properties and play an important role in various processes. In order to prevent overlapping particles at the beginning of the simulations, the rectangular box was divided into equal-sized basic cells. Each triplet was then placed into a basic cell with a randomly assigned orientation. Both translational and rotational movements of each triplet were taken into account during the simulations. The translational equations of motion were numerically solved using the standard leapfrog method~\cite{Hockney1970, Potter1973}, while the rotational equations were handled using an implicit rotational leapfrog scheme in the quaternion representation (see section \ref{sec:rls}). The simulations considered both contact and long-range cohesive forces. The contact forces arise from the deformation of the primary spheres that make up the colliding triplets and can be classified into two main types: normal viscoelastic force and tangential force. The normal viscoelastic force was calculated using the nonlinear Hertz model~\cite{Popov2010, Brilliantov1996}, whereas the tangential force was determined using a nonlinear-spring model derived from the Mindlin--Deresiewicz theory~\cite{Mindlin1953}. 

To incorporate long-range forces, a modified LJ potential~\cite{Ferraz2018, Ferraz2019} based on the molecular framework assumption was employed. The packing processes were investigated by varying the strength of long-range interactions. Statistical calculations were performed to analyze various quantities throughout the system's evolution, including the kinetic-to-potential energy ratio, packing density, and mean coordination number. In addition, radial distribution and orientation pair correlation functions were computed to characterize the microstructure of the formed packs.

The manuscript is structured as follows. Section \ref{sec:mms} provides a detailed description of the model and DEM simulations employed. Section \ref{sec:r} presents and discusses the obtained results. Finally, in Section \ref{sec:c}, the conclusions drawn from the study are presented.

\section{Model and DEM simulations \label{sec:mms}}

\begin{figure*}[!t]
\centering
\begin{minipage}[t]{1.0\linewidth}
\centering
\subfigure[]{\label{fig:02a}\includegraphics[scale=0.58, angle=0]{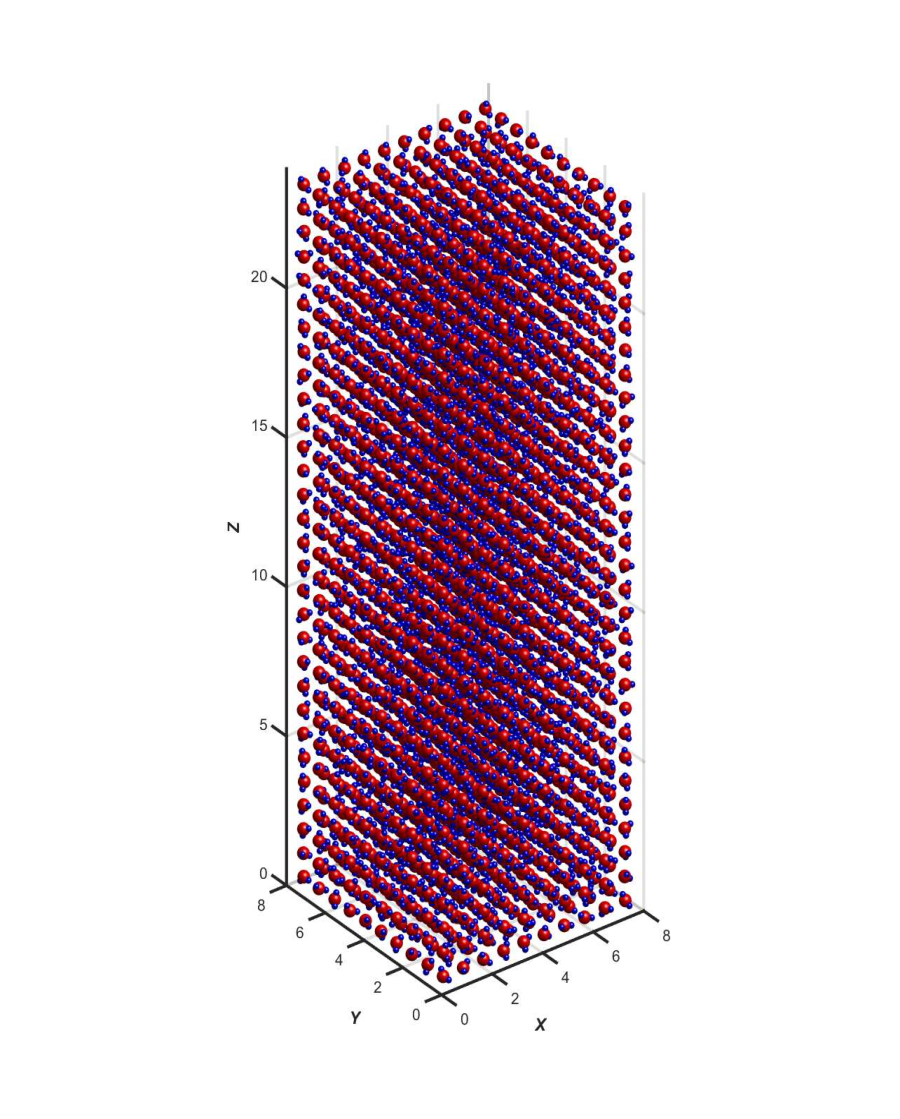}}
\subfigure[]{\label{fig:02b}\includegraphics[scale=0.48, angle=0]{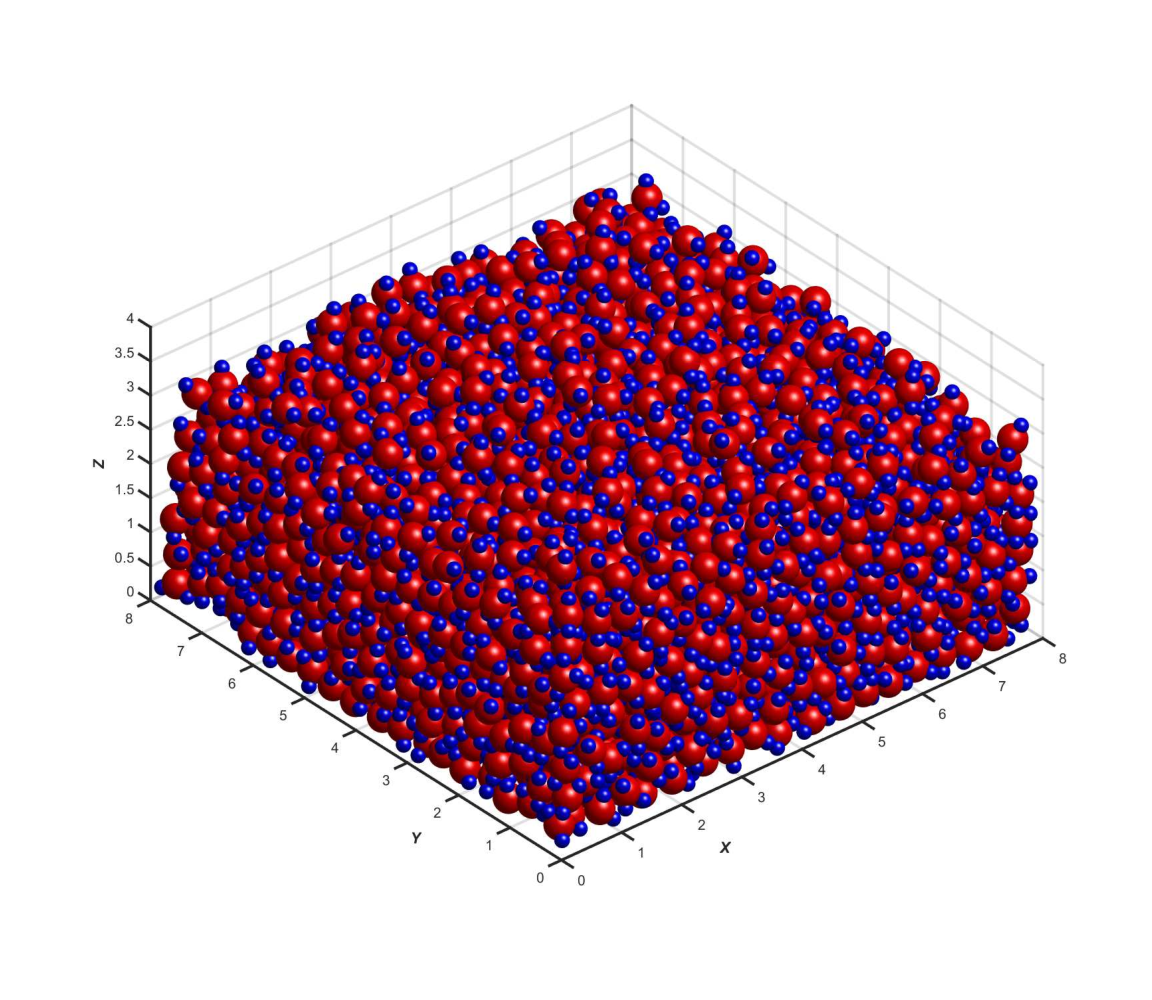}}
\end{minipage}
\caption{Snapshots of a typical packing process with $\Sigma_{2v}(2\pi/3)$-triplet without long-range interaction ($\varepsilon=0$) inside a $(8 \times 8 \times 24) \, \mu m$ box at the instants (a) $t=0.0 \, ms$ and (b) $t=5.0 \, ms$. Fig.~\ref{fig:02b} provides a close-up view of the formed particle aggregate.The parameters used in this simulation are given in Table \ref{table:01}.}\label{fig:02}
\end{figure*}

\subsection{Equations of motion}
 The equations of motion for a non-spherical particle $i$ of mass $m_{i}$ and inertial tensor $\mathbf{I}_{i}$, are given by

\begin{equation} \label{eq:1}
{m_i}\frac{{d{\kern 1pt} {{\vec v}_i}}}{{dt}} = {\vec F_i} + {m_i}\vec g
\end{equation}
and 
\begin{equation} \label{eq:2}
 {{\mathbf I}_i}\frac{{d{{\vec \omega }_i}}}{{dt}} + {\vec \omega _i} \times ({{\mathbf I}_i}{\vec \omega _i}) = {\vec \tau _i}.
\end{equation}
Eqs.(\ref{eq:1}) and (\ref{eq:2}) rule the translational motion of the particle's center of mass and rotation motion of the particle about its center of mass in the body-fixed coordinate frame, respectively. $\vec v_{i}$ and $\vec \omega_{i}$ are the corresponding linear and angular velocities of the particle, and $\vec g$ is the gravity acceleration. $\vec F_{i}$ and $\vec \tau_{i}$ are the sum of all external forces and torques acting on the particle $i$, respectively.

In the present model, the resulting force $\vec F_{i}$ is due to interactions between different composite particles (triplets) and can be expressed by
\begin{equation} \label{eq:3}
{\vec F_i} = \sum\limits_{j = 1}^M {(\sum\limits_{\alpha  = 1}^{{N_i}} {\sum\limits_{\beta  = 1}^{{N_j}} {(\vec F_{\alpha \beta }^n\, + \vec F_{\alpha \beta }^t)\,} } } \theta ({\delta _n})) + \sum\limits_{j = 1}^{MM} {F_{ij}^{LJ}} ,
\end{equation}
where $\vec F_{\alpha \beta}^n$ is the normal viscoelastic force and $\vec F_{\alpha \beta}^t$ is the tangential friction force between the $\alpha$- and $\beta$-th primary spheres, while $\vec F_{ij}^{LJ}$ is the LJ force between the $i$- and $j$-th composite particles. In Eq.~(\ref{eq:3}), $\theta ({\delta _n})$ is the Heaviside function
\begin{equation} \label{eq:4}
	\theta ({\delta _n}) =
	\begin{cases}
		1&   \text{if $\delta_n>0$}, \\
		0&   \text{if $\delta_n\leq 0$},
	\end{cases}
\end{equation}
being $\delta_{n}$ the normal deformation (see Eq.~(\ref{eq:6})). In the summations above, $M$ represents the number of collisions that the $i$-th particle undergoes with $j$-th others, whilst $MM$ represents the number of long-range interactions between the $i$-th particle and $j$-th others at each time step. While $N_{i}$ and $N_{j}$ denote the number of colliding spheres that compose the $i$- and $j$-th particle, respectively. 

The normal viscoelastic force $\vec F_{\alpha\beta}^n$ is derived from the nonlinear Hertz theory, and it can written as
\begin{equation} \label{eq:5}
 \vec F_{\alpha\beta}^n = [\frac{2}{3}E\sqrt {\bar R}\, \delta _{n}^{3/2}-{\gamma _n}E\sqrt {\bar R} \sqrt {{\delta _{n}}} ({{\vec v}_{\alpha\beta}} \cdot {{\hat n}_{\alpha\beta}})]{{\hat n}_{\alpha\beta}},
\end{equation}
where $E=Y/(1-\xi^2)$ is the elastic modulus of two primary spheres, being $Y$ the Young's modulus and $\xi$ the Poisson' ratio. $\bar R=R_{\alpha}R_{\beta}/(R_{\alpha}+R_{\beta})$ is the effective radius, $\delta _{n}$ is the deformation which is expressed by
\begin{equation} \label{eq:6}
	\delta _{n}=(R_{\alpha}+R_{\beta})-(|\vec{r}_{\alpha}(t)-\vec{r}_{\beta}(t)|),
\end{equation}
 $\vec{v}_{\alpha\beta}$ is the relative velocity between $\alpha$- and $\beta$-th primary spheres, and $\gamma _{n}$ is the normal damping  coefficient. 

The tangential friction force $\vec F_{\alpha\beta}^t$ is calculated according to the Mindlin-Deresiewicz theory as
\begin{equation} \label{eq:7}
\vec F_{\alpha\beta}^t = {\gamma _t}|\vec F_{\alpha\beta}^n|\left[ {1 - {{\left( {1 - \frac{{|{\delta _{t}}|}}{{|{\delta _{\max }}|}}} \right)}^{3/2}}} \right]{{\hat t}_{\alpha\beta}},
\end{equation}
where  $\gamma _t$ is the friction coefficient, ${\hat t}_{\alpha\beta}$ is the unit vector perpendicular to ${\hat n}_{\alpha\beta}$, $\delta _{t}$ is the tangential displacement which is determined as
\begin{equation} \label{eq:8}
{\delta _{t}} = \int\limits_{{0}}^{t_{c}} {({{\vec v}_{\alpha\beta}} \cdot {\kern 1pt} {\kern 1pt} {\kern 1pt} } {{\hat t}_{\alpha\beta}} + {R_\alpha}\,{{\hat n}_{\alpha\beta}} \times {{\vec \omega }_\alpha} + {R_\beta}\,{{\hat n}_{\alpha\beta}} \times {{\vec \omega }_\beta})dt,
\end{equation}
where the above integral is calculated during the contact time $t_{c}$ between the particles. The $\delta _{\max }$ is the maximum tangential displacement and in the condition that $\delta _{t}>|\delta _{\max}|$, the sliding friction takes place between the particles.

To optimize computational efficiency, a coarse-grained Lennard-Jones (LJ) potential was employed to compute the cohesive long-range forces acting between two composite particles $i$ and $j$. This was achieved by introducing an effective radius, $R_{eff}$, for each particle. The expression for the LJ force, $\vec F_{ij}^{LJ}$, can be written as follows:
\begin{equation} \label{eq:9}
\vec F_{ij}^{LJ} = \sum\limits_{k = 1}^m \frac{{24\,\varepsilon}}{\sigma }\left[ {2{{\left( {\frac{\sigma}{{{r_{ij}}}}} \right)}^{13}} - {{\left( {\frac{\sigma}{{{r_{ij}}}}} \right)}^7}} \right]{{\hat n}_{ij}},
\end{equation} 
where $r_{ij}$ is the distance between the centers of mass of the particles, and $m$ in the sum is the number of primary spheres that make up the particles. Here $\varepsilon$ is an important control parameter, which rules the strength of the interaction, and $\sigma=2^{-5/6}R_{eff}$ defines the hard core of the potential. It is important to say that the LJ force is only activated when $r_{ij}>2R_{eff}$. When $r_{ij}\leq 2 \,R_{eff}$, the dominant forces governing the motion of the particles are the contact forces described by Eq.~(\ref{eq:6}) and (\ref{eq:7}). Furthermore, we have defined a cutoff at $r_{ij}=3\,R_{eff}$, which was utilized with the linked-list method~\cite{Allen1989, Rapaport1995} to save time during the simulations. The average CPU time to update the state of one particle was about $5.68\, ns$ on a 3.10 GHz Intel PC.

\subsection{Rotation and the quaternion representation}

\begin{table}[!b]
\centering
\small
\begin{threeparttable}
\caption{\label{table:01} Physical parameters used in the simulations.}
\begin{tabular}{lc}
\hline \hline \\
  Parameter\tnote{a} &  Value \\ \hline \\
  Number of particles ($N$) & 3000 \\
  Particle effective radius ($R_{eff}$),& $3.15 \times 10^{-7} m$ \\
  Particle density ($\rho$) & $1672.91\;$ $kg/m^{3}$ \\
	Range of minimum potential energy ($\varepsilon$) & $0-65.0\,\mu J$\\
  Young's modulus ($Y$) &  $10^{8}\,$ $N/m^{2}$ \\
	Poisson's ratio ($\xi$) & 0.30 \\
	Normal damping coefficient ($\gamma_{n}$) &  $5.0\times 10^{-4} s$  \\
  Tangential damping coefficient ($\gamma_{t}$) &  0.30 \\
\hline \hline
\end{tabular}
\begin{tablenotes}
      \item[a]{It is assumed that both particles and walls have the same physical parameters.}
    \end{tablenotes}
\end{threeparttable}
\end{table}

\begin{figure}[!t]
	\centering
	\includegraphics[scale=0.37, angle=0]{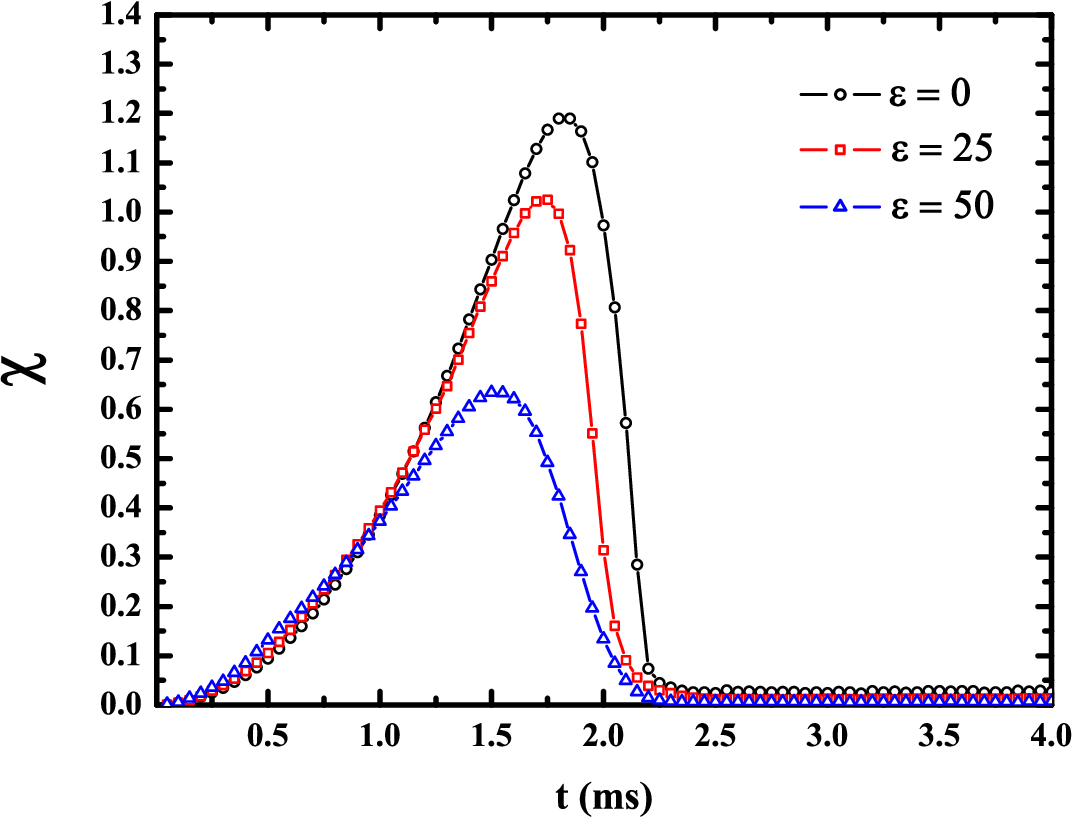}
\caption{Plot of the average ratio $\chi$ of the kinetic energy ($K$) to the gravitational potential energy ($U$) for particles $\Sigma_{2v}(2\pi/3)$ as a function of time, considering three different $\varepsilon$ values.}\label{fig:03}
\end{figure}

The orientation of a body in space can be specified by considering the relationship between the space-fixed coordinate system and the body-fixed coordinate system~\cite{Goldstein2002}. Both system are defined such that their origins coincide when the particle is not rotated. A vector $\vec u$ can be expressed in terms of the components in both systems. We will use the notation $\vec {u}^{s}$ for the representation in the space-fixed system and $\vec u^{b}$ for the representation in the body-fixed system. The components of the vector in both frames are related by the equations
\begin{equation} \label{eq:10}
	{\vec u^b} = A \cdot {\vec u^s}
\end{equation}
and
\begin{equation} \label{eq:11}
	{\vec u^s} = {A^{T}} \cdot {\vec u^b},
\end{equation}
where $A$ is the $3\times3$ rotation matrix such that $A^{T}=A^{-1}$. However, an active rotation takes place when the vector $\vec u$ itself rotates in a fixed coordinate system, i.e.,
\begin{equation} \label{eq:12}
	\vec u' = {A^T} \cdot \vec u,
\end{equation}
being therefore equivalent to Eq.~(\ref{eq:11}). Moreover, if the vector $\vec u$ is embedded in the body frame, it is easy to show that
\begin{equation} \label{eq:13}
\vec {\dot u}^{s} = \vec \omega^{s}  \times \vec u^{s},
\end{equation}
yielding equations of motion for any three orthogonal vectors (or $A$ rows) that orient the body in the space-fixed coordinate system. Also, we can write Eq.~(\ref{eq:2}), dropping the index $i$, in their respective principal-axis components 
\begin{equation} \label{eq:14}
	\begin{gathered}
  \dot \omega _x^b = \frac{{\tau _x^b}}{{I_{xx}^b}} + \left( {\frac{{I_{yy}^b - I_{zz}^b}}{{I_{xx}^b}}} \right)\,\omega _y^b\omega _z^b, \hfill \\
  \dot \omega _y^b = \frac{{\tau _y^b}}{{I_{yy}^b}} + \left( {\frac{{I_{zz}^b - I_{xx}^b}}{{I_{yy}^b}}} \right)\,\omega _x^b\omega _z^b, \hfill \\
  \dot \omega _z^b = \frac{{\tau _z^b}}{{I_{zz}^b}} + \left( {\frac{{I_{xx}^b - I_{yy}^b}}{{I_{zz}^b}}} \right)\,\omega _x^b\omega _y^b. \hfill \\ 
\end{gathered}
\end{equation}
Eqs.~(\ref{eq:13}) and (\ref{eq:14}) provide the basic framework for determining the body orientation over time. Yet it is necessary to define a representation for the orientation. Since Eqs.~(\ref{eq:10}) and (\ref{eq:11}) will have to be extensively used, it is helpful to obtain a parametric representation for $A$. 

Only three independent quantities are necessary to define a $3\times 3$ rotation matrix $A$. Euler angles~\cite{Goldstein2002, Trindade2001} are a popular choice in many situations. However, equations of motion using Euler angles are susceptible to singularities. As a result, alternative parameterizations have been used in DEM, including Cayley-Klein parameters~\cite{Goldstein2002}, conformal rotation vector~\cite{Milenkovic82}, and quaternions~\cite{Evans1977, Kuipers2002}. In this work, we will utilize quaternions for the parameterization of the matrix $A$.

A quaternion $\mathbf q$ can be defined as the complex sum of a scalar $q_0$ and a vector $\vec q=(q_1, q_2, q_3)$, that is, ${\mathbf q} = {q_0} + i{\kern 1pt} \vec q$. Here we use the notation,
\begin{equation} \label{eq:15}
{\mathbf q} = ({q_0},\vec q) \equiv ({q_0},{q_1},{q_2},{q_3}).
\end{equation}
A quaternion can be interpreted as representing a rotation around an axis defined by a unit vector $\hat n$, by an angle $\varphi$ as
\begin{equation} \label{eq:16}
\mathbf{q} = (\cos \frac{\varphi }{2},\,\hat n\sin \frac{\varphi }{2}).
\end{equation}
While the product of two quaternions $\mathbf{q}$ and $\mathbf{p}$ is define as
\begin{equation}\label{eq:17}
\begin{split}
\mathbf{q} \otimes \mathbf{p} &= ({q_0}{p_0} - \vec q \cdot \vec p,\,\;{q_0}\vec p + {p_0}\vec q + \vec q \times \vec p) \\
 &= \begin{pmatrix}
  {{q_0}}&{ - {q_1}}&{ - {q_2}}&{ - {q_3}} \\ 
  {{q_1}}&{{q_0}}&{ - {q_3}}&{{q_2}} \\ 
  {{q_2}}&{{q_3}}&{{q_0}}&{ - {q_1}} \\ 
  {{q_3}}&{ - {q_2}}&{{q_1}}&{{q_0}} 
\end{pmatrix}\; \begin{pmatrix}
  {{p_0}} \\ 
  {{p_1}} \\ 
  {{p_2}} \\ 
  {{p_3}} 
\end{pmatrix}. \\
\end{split}
\end{equation}
Through Eqs.~\eqref{eq:16} and \eqref{eq:17}, a rotation given by Eq.~\eqref{eq:12} can be rewritten as 
\begin{equation} \label{eq:18}
\vec {u'} = \mathbf{q} \otimes \vec{u} \otimes {\mathbf{q^{ - 1}}},
\end{equation}
where ${\mathbf{q^{ - 1}}}=({q_{0}},-\vec q) $, and we have also defined a quaternion $\mathbf{u}=(0,\vec u)$.
Using equations \eqref{eq:12}, \eqref{eq:16}, and \eqref{eq:17}, we can determine the rotation matrix in quaternion representation, which can be expressed by
\begin{equation} \label{eq:19}
A=\begin{pmatrix}
  {q_0^2 + q_1^2 - q_2^2 - q_3^2}&{2({q_1}{q_2} + {q_0}{q_3})}&{2({q_1}{q_3} - {q_0}{q_2})} \\ 
  {2({q_1}{q_2} - {q_0}{q_3})}&{q_0^2 - q_1^2 + q_2^2 - q_3^2}&{2({q_2}{q_3} + {q_0}{q_1})} \\ 
  {2({q_1}{q_3} + {q_0}{q_2})}&{2({q_2}{q_3} - {q_0}{q_1})}&{q_0^2 - q_1^2 - q_2^2 + q_3^2} 
\end{pmatrix}.
\end{equation}

It can be demonstrated~\cite{Powles1979,Sonnenschein1985,Rapaport1985} that the first and second derivatives of the quaternion $\mathbf{q}$, which describe the particle's orientation, are given by:
\begin{equation} \label{eq:20}
\mathbf{\dot q} = \frac{1}{2}\mathbf{q} \otimes \mathbf{w^b}
\end{equation}
and
\begin{equation}\label{eq:21}
\mathbf{\ddot q} = \frac{1}{2}\mathbf{q} \otimes \mathbf{\dot W^b},
\end{equation} 
where the above quaternions $\mathbf{w^b}$ and $\mathbf{\dot W^b}$ are defined by
$\mathbf{w^b}=(0,\omega_{x}^{b},\omega_{y}^{b},\omega_{y}^{b})$ and $\mathbf{\dot W^b}=(-2\sum \dot q_m^2,\dot \omega_{x}^{b},\dot \omega_{y}^{b},\dot \omega_{y}^{b})$.

\subsection{Rotational leapfrog scheme using quaternion acceleration \label{sec:rls}}

\begin{figure*}[!t]
	\centering
	\begin{minipage} [t]{0.49\linewidth}
\centering
	\includegraphics*[scale=0.37, angle=0]{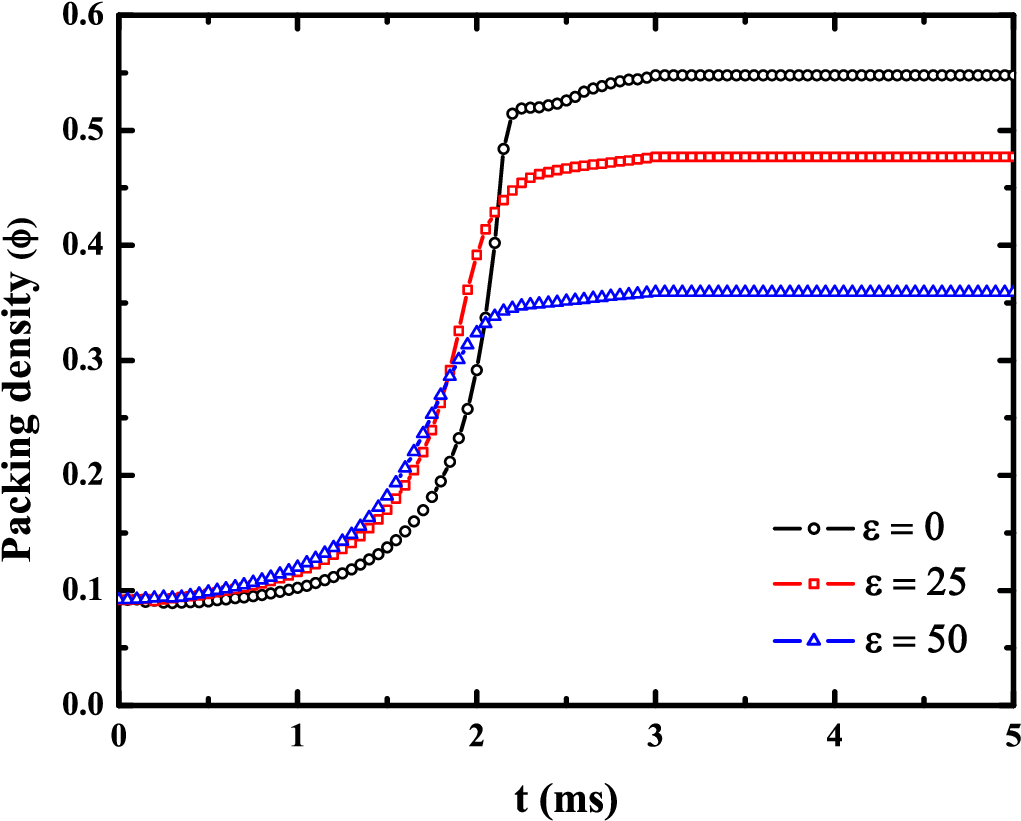}
\caption{Plot of the packing densities $\phi$ for $\Sigma_{2v}(2\pi/3)$-triplets as a function of time, considering three different long-range interaction strengths $\varepsilon$. Data points are averages over $10$ independent realizations.}\label{fig:04}
\end{minipage}\hfill
\begin{minipage}[t]{0.49\linewidth}
	\centering
	\includegraphics[scale=0.37, angle=0]{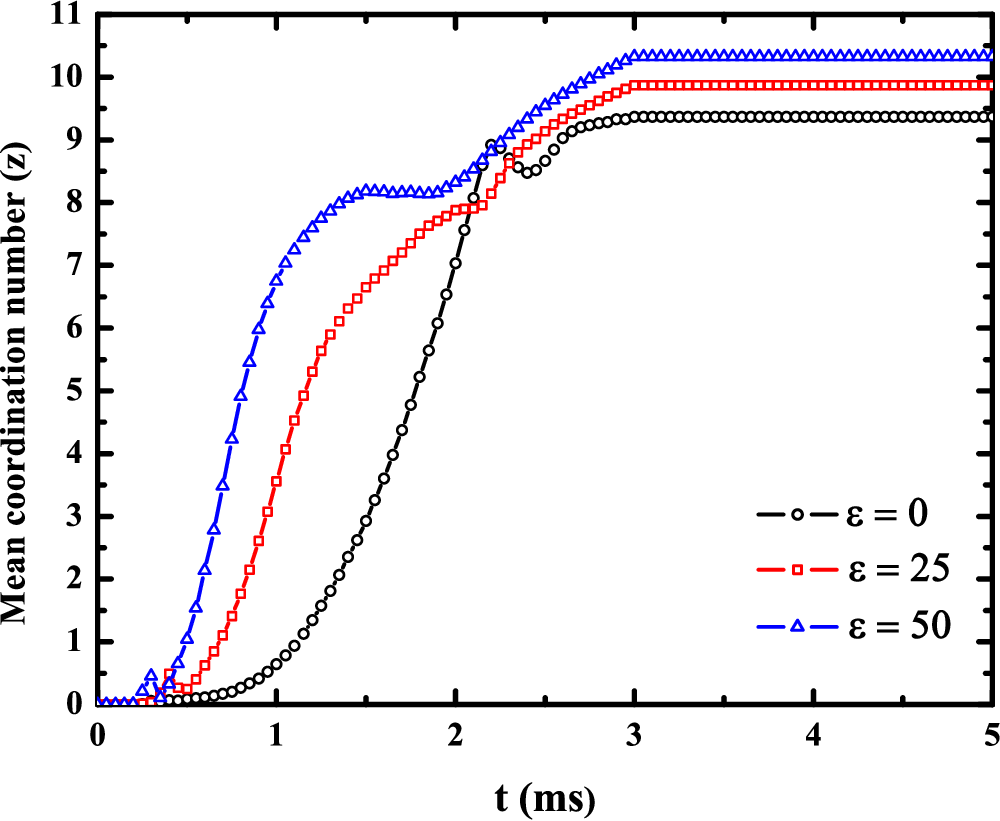}
\caption{Plot of the mean coordination number $z$ for $\Sigma_{2v}(2\pi/3)$-triplets as a function of time, considering three different long-range interaction strengths $\varepsilon$. Data points are averages over $10$ independent realizations.}\label{fig:05}
\end{minipage}
\end{figure*}

\begin{figure*}[t]
\centering
\begin{minipage} [t]{0.49\linewidth}
\centering
\includegraphics*[scale=0.37]{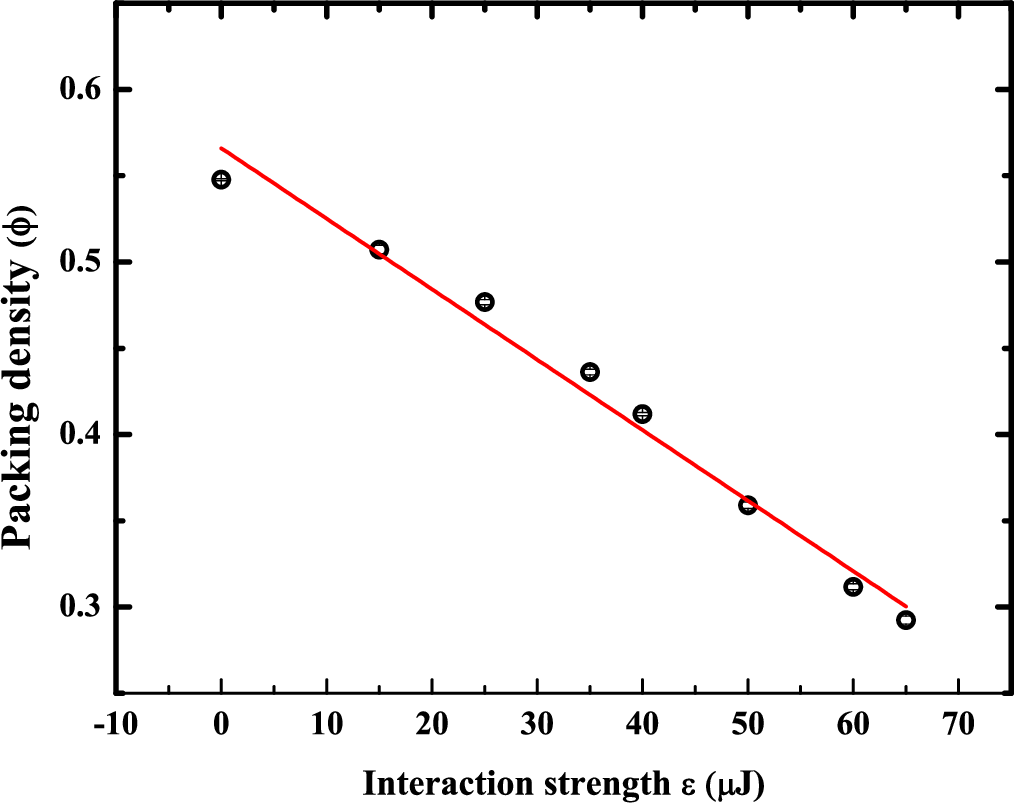}
 \caption{Plot of the packing densities $\phi$ for $\Sigma_{2v}(2\pi/3)$-triplets as a function of the interaction strength $\varepsilon$. The red straight line is the best linear fit to data points. Error bars were calculated by averaging 10 independent runs and are smaller than the size of the plotting symbols.} \label{fig:06}
\end{minipage}\hfill
\begin{minipage}[t]{0.49\linewidth}
\centering
\includegraphics[scale=0.37]{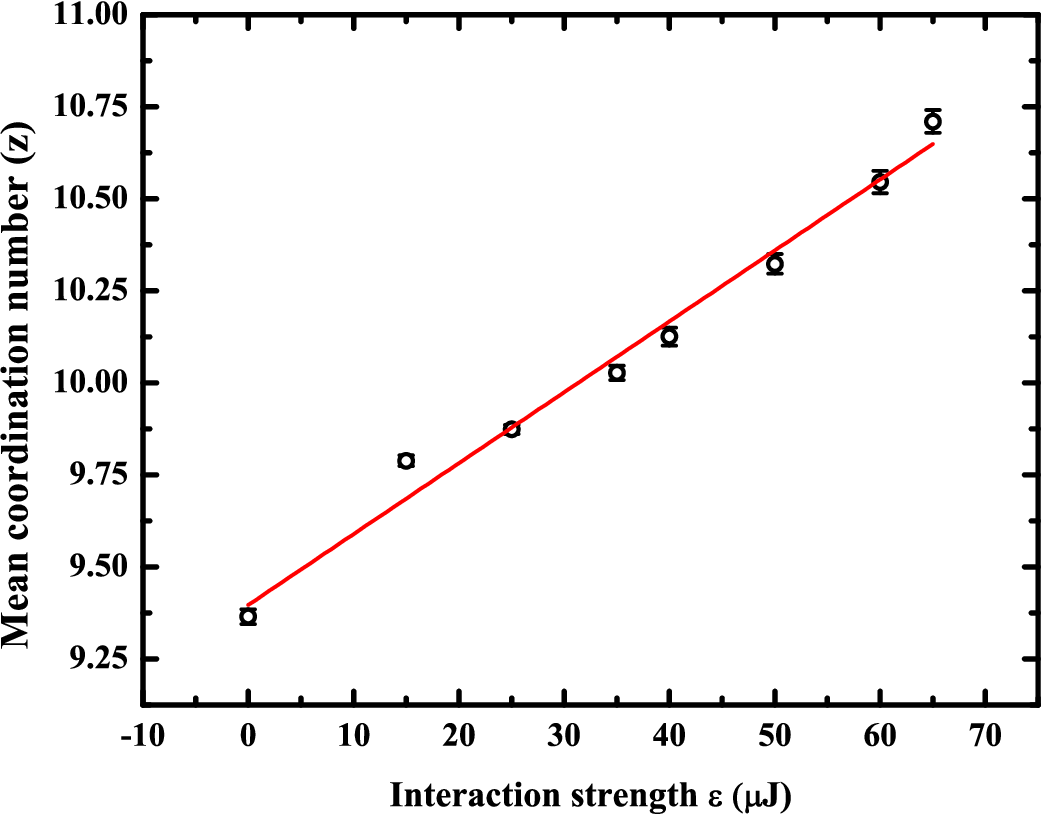}
 \caption{Plot of the  mean coordination number $z$ for $\Sigma_{2v}(2\pi/3)$-triplets as a function of the interaction strengths $\varepsilon$. Likewise, line and error bars are as in Fig.~\ref{fig:06}.} \label{fig:07}
\end{minipage}
\end{figure*}

\begin{figure*}[t]
\centering
\begin{minipage} [t]{0.49\linewidth}
\includegraphics*[scale=0.36]{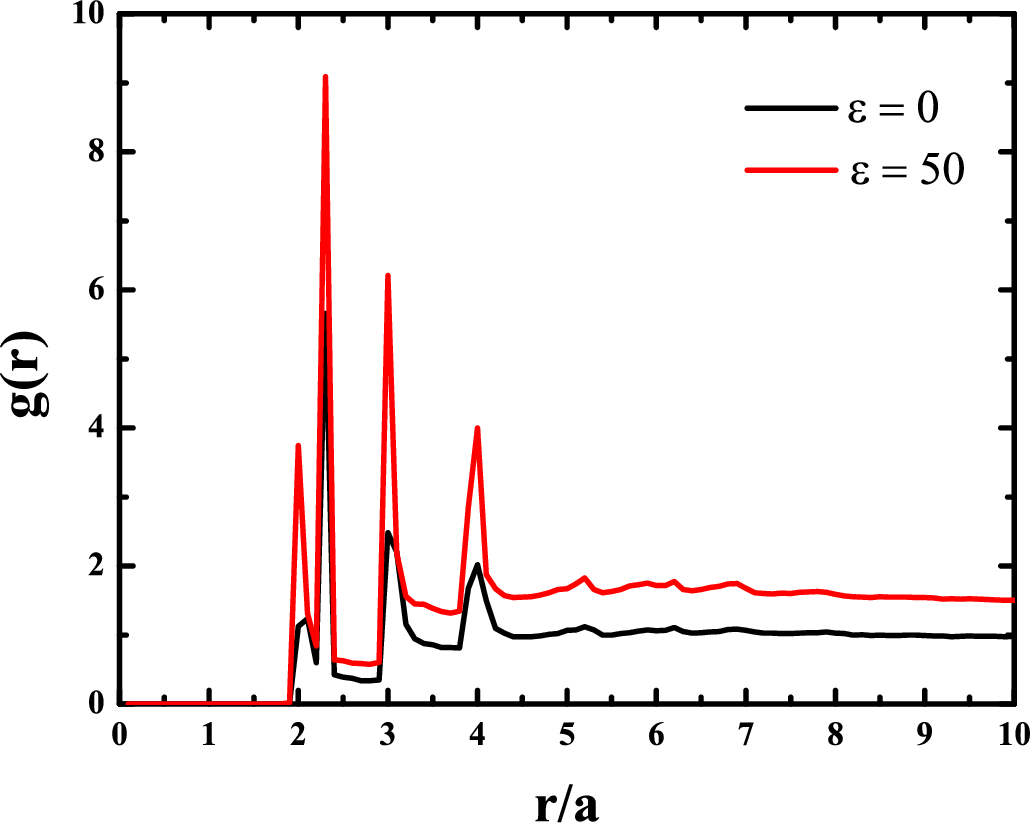}
\centering
 \caption{RDFs of the random packing structures formed by $\Sigma_{2v}(2\pi/3)$-triplets, considering two different $\varepsilon$ values ($a=0.10\, \mu m$). The black line represents RDF when $\varepsilon=0\, \mu J$ (absence of long-range interactions) and the red line represents RDF when $\varepsilon=50\, \mu J$.} \label{fig:08}
\end{minipage}\hfill
\begin{minipage}[t]{0.49\linewidth}
\includegraphics[scale=0.36]{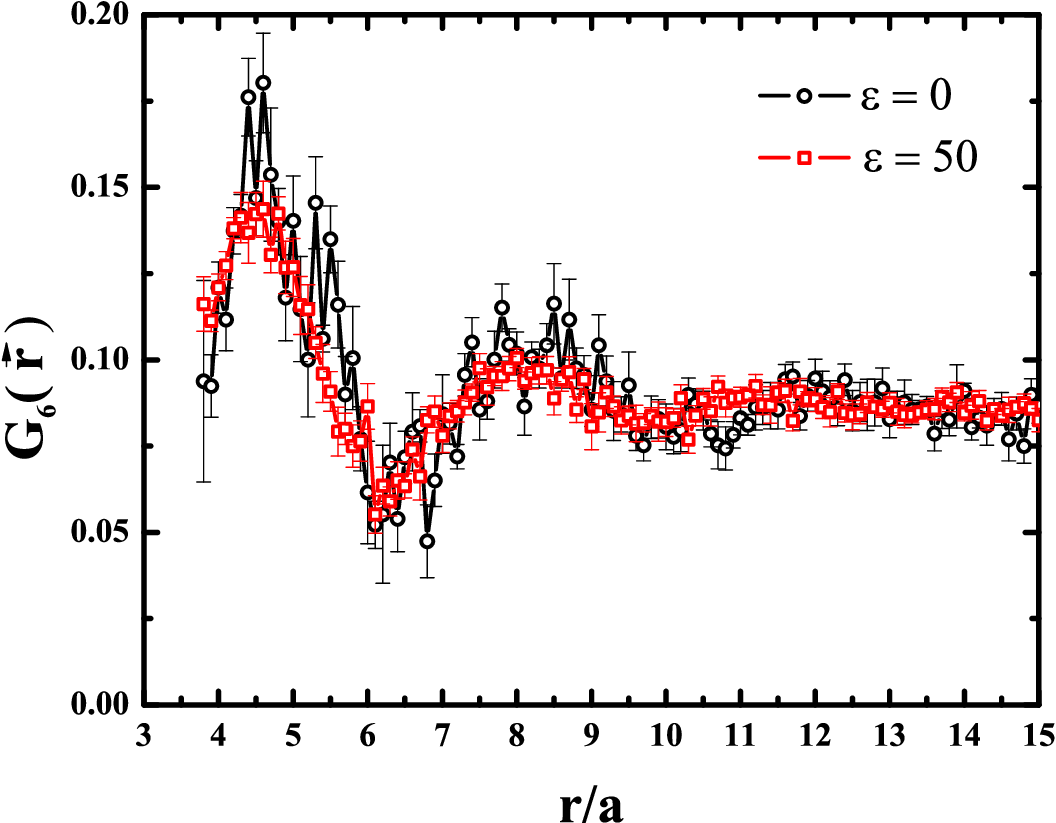}
\centering
 \caption{Plot of the orientation pair correlation function $G_{6}$ as a function of the radial distance for random packing structures formed by $\Sigma_{2v}(2\pi/3)$-triplets, considering two different $\varepsilon$ values ($a=0.10\, \mu$m). The black line represents $G_{6}$ when $\varepsilon=0\, \mu J$  and the red line represents $G_{6}$ when $\varepsilon=50\, \mu J$. Error bars were calculated by averaging 10 independent realizations. }\label{fig:09}
\end{minipage}
\end{figure*}

\begin{figure*}[t]
\centering
\begin{minipage} [t]{0.49\linewidth}
\includegraphics*[scale=0.36]{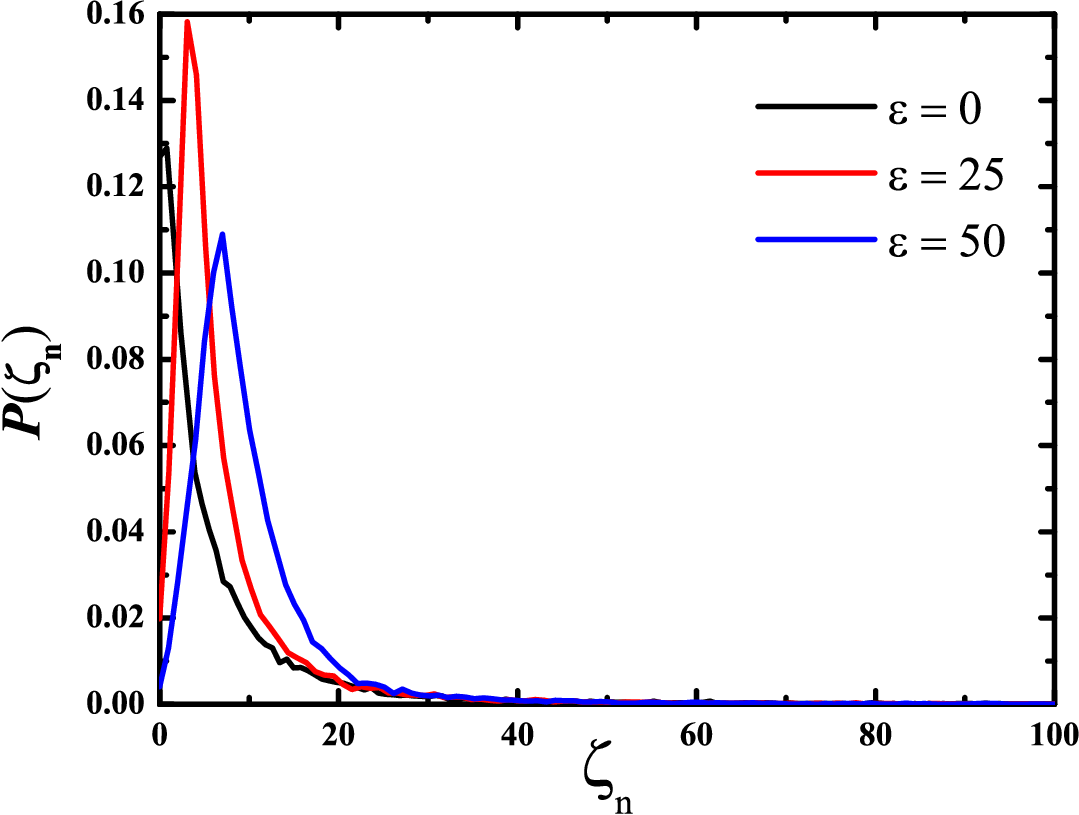}
\centering
 \caption{Normal contact force distribution in the random packing structures with $\Sigma_{2v}(2\pi/3)$-triplets by considering three different $\varepsilon$ values. $\zeta _{n}$ is the ratio of the magnitudes of the normal contact force to the magnitude of the gravitational force.} \label{fig:10}
\end{minipage}\hfill
\begin{minipage}[t]{0.49\linewidth}
\includegraphics[scale=0.36]{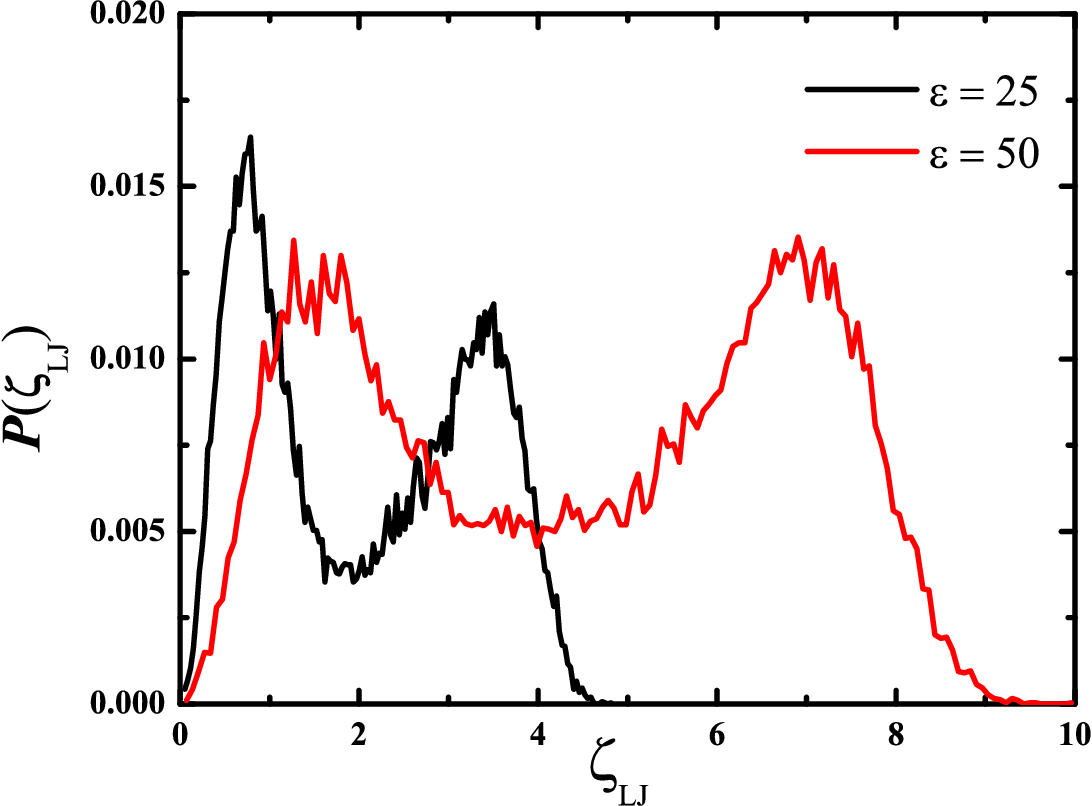}
\centering
 \caption{Long-range force (LJ force) distribution in the random packing structures with $\Sigma_{2v}(2\pi/3)$-triplets by considering two different $\varepsilon$ values. $\zeta _{LJ}$ is the ratio of the magnitudes of the LJ force to the magnitude of the gravitational force.}\label{fig:11}
\end{minipage}
\end{figure*}

For each particle $i$, Eq.~\eqref{eq:20} can be numerically integrated through an implicit leapfrog scheme, which is outlined as follows: 
\begin{enumerate}
\item Starting from the knowledge of the quantities $\mathbf{q}_{i}(n)$, $\mathbf{\dot q}_{i}(n-1/2)$, $\vec{\tau_{i}}^{b}(n)$ and principal moment of inertia $\mathbf{I}_{i}^{b}$ at the $n$-th time step, obtain a first estimate for $\vec{\omega_{i}}^{b}(n)$ using $\mathbf{\dot q}_{i}(n-1/2)$ and the inverse of Eq.~\eqref{eq:20}. 
\item Use the results for $\vec{\omega_{i}}^{b}(n)$ to calculate $\mathbf{\ddot q}_{i}(n)$ from Eq.~\eqref{eq:21} and propagate the quaternion velocity $\mathbf{\dot q}_{i}$ to one step further
\begin{equation}\label{eq:22}
 \mathbf{\dot q}_{i}(n+1/2)=\mathbf{\dot q}_{i}(n-1/2)+\Delta t\,\mathbf{\ddot q}_{i}(n)+\mathscr{O}(\Delta t^{3}).
 \end{equation} 
 \item Recalculate $\mathbf{\dot q}_{i}(n)$ using the following interpolation:
 \begin{equation} \label{eq:23}
 \mathbf{\dot q}_{i}(n)=\dfrac{1}{2}\big(\mathbf{\dot q}_{i}(n+1/2)+\mathbf{\dot q}_{i}(n-1/2)\big).
 \end{equation}
 \item Use $\mathbf{\dot q}_{i}(n)$ from step $3$ in Eq.~\eqref{eq:20} to get an improved estimate of $\vec{\omega_{i}}^{b}(n)$ and then recalculate $\mathbf{\ddot q}_{i}(n)$ through Eq.~\eqref{eq:21}. 
 \item Repeat steps 2--4 until $\vec{\omega_{i}}^{b}(n)$ and $\mathbf{\dot q}_{i}(n+1/2)$ converge.
 \item Advance $\mathbf{q}_{i}$  by one time-step in a leapfrog fashion:
 \begin{equation} \label{eq:24}
 \mathbf{q}_{i}(n+1)=\mathbf{q}_{i}(n)+\Delta t\,\mathbf{\dot q}_{i}(n+1/2)+\mathscr{O}(\Delta t^{4}).
 \end{equation}
 \item Calculate the new rotation matrix $A_{i}^{T}$ from $\mathbf{q}_{i}(n+1)$ using Eq. \eqref{eq:19} and rotate the $i$-th particle to its new orientation at step $n+1$ via Eq.~\eqref{eq:12}.
 \item Repeat all the above-mentioned steps for the subsequent time steps. 
\end{enumerate}
The validity of the order of error in Eqs.~\eqref{eq:22} and \eqref{eq:24} is established only upon reaching convergence in the iterations. It can be observed that a few iterations are sufficient to achieve the desired accuracy level in step $5$.

\section{\label{sec:r} Results and Discussion}

In this work, the particle packing processes were investigated using particles called $\Sigma_{2v}(2\pi/3)$-triplets. Each triplet is formed by overlapping two primary spheres of radius $a=0.10 \, \mu m$ with one sphere of radius $b=0.20 \, \mu m$, arranged in an angular geometry as shown in Fig.~\ref{fig:01}. Considering the overlap regions between adjoining primary spheres, the total volume $V$ of each triplet is given by
\begin{equation} \label{eq:22}
\begin{split}
V &=\frac{4 \pi}{3}(2{a^3} + {b^3}) \\
&-\pi \frac{{{{(a + b - d)}^2}({d^2} + 2ad - 3{a^2} + 2bd + 6ab - 3{b^2})}}{{6d}} \\
& \approx 39.85 \times 10^{-21} m^{3} ,\\
\end{split}
\end{equation}
being $d = \frac{3}{4}(a + b)$ the distance between the centers of two adjoining spheres. Three thousand of particles were introduced into a micrometer-sized ($8\, \times 8\, \times 24$) box. To overcome the challenging issue of inserting non-overlapping particles into the box, a solution was implemented by dividing it into basic cubic cells [$(a_{0},a_{0},a_{0})$, $a_{0}=2(b+2a)$] capable of hosting a single particle inside. Each particle was initially assigned random Euler angles by using a random number generator~\cite{recipes96}. Moreover, in order to avoid the complicating effects of the pouring rate, the particles were suspended along the box at the beginning of the simulation. After that, the particles were pulled down by gravity and started to collide each other. Both particle-particle and particle-wall interactions were taken into account in the simulations.
	
	The packing process for triplets without long-range interaction ($\varepsilon=0$) is illustrated in the Fig.~\ref{fig:02}. Snapshots at the instants $t=0\, ms$ (Fig.~\ref{fig:02a}) and $t=5.0 \, ms$ (Fig.~\ref{fig:02a}) are shown in this figure. Note that Fig.~\ref{fig:02b} presents a close-up view of the particle aggregate that has been formed. The vertical position of the uppermost surface of the box is determined by calculating the average height (z-coordinate) of the centers of mass of the particles constituting the top layer of the particle assembly. In order to mitigate finite-size box effects, all packing quantities were calculated inside a smaller virtual box with virtual walls positioned at least $0.30 \, \mu m$ away from the actual walls of the confinement box. We performed statistical calculations on various quantities, including packing density and mean coordination number as the system evolved over time.  To obtain the average values of these quantities and estimate their statistical error, we averaged over $10$ independent realizations. The physical parameters used in the simulations are given in Table \ref{table:01}.
	
	In Fig.~\ref{fig:03} is shown the time evolution of the average ratio $\chi=K/U$ of the kinetic energy ($K$) to the gravitational potential energy ($U$) for three different $\varepsilon$ values. Similar energy curves were found for other remaining $\varepsilon$ values. From this figure, one can see that the system relaxation was already achieved after $2.50\, ms$ for all $\varepsilon$ values considered. The slightly longer relaxation time for the case $\varepsilon=0$ is mainly due to the higher degree of freedom of the particles, as they bounce more times after hitting the bottom of the box before being halted by dissipative forces. This also explains the higher peak observed in the energy ratio curve for this case. Furthermore, it is evident that the energy peak decreases as the values of $\varepsilon$ increase.
	
	The packing densities $\phi$ for $\Sigma_{2v}(2\pi/3)$-triplets are shown as a function of time in Fig.~\ref{fig:04}, considering several long-range interaction strengths $\varepsilon$.  In this figure, the $\phi$ values are given at short time intervals of $50.0\, \mu s$ up to  $5\, ms$. After relaxation time the ultimate $\phi$ values were obtained for each case. The initial packing densities were about $9.20 \times 10^{-2}$ for all cases. Each data point shown represents an average calculated from 10 independent realizations, obtained by varying the initial orientation of the particles. The packing density minimum around $2.5 \,ms$ for the case $\varepsilon=0$ was due to the first particles' bouncing after hitting the bottom base of the box. As expected, the packing density of the formed packs decreases with increasing interaction strength $\varepsilon$. This is mainly due to the additional space created by long-range forces between the virtual walls and the outermost layers of the formed samples. In fact, the larger the $\varepsilon$ value, the greater the final height of the particle stack, resulting in increased space between the virtual walls and the particle aggregate. The final $\phi$ values obtained were as follows: 0.54(8) for the $\varepsilon=0$ case, 0.47(7) for the $\varepsilon=25\,\mu J$ case, and 0.35(9) for the $\varepsilon=50\,\mu J$ case. In all cases, the $\phi$ values were significantly lower than $\pi/\sqrt{18}\simeq 0.74$ as reported in the literature~\cite{Zamponi2008}, which corresponds to closest-packing crystal structures, namely, face-centered cubic (fcc) and hexagonal close-packed (hcp) structures. Interestingly, the $\phi$ values for all the cases studied here were found to be below $\phi\simeq 0.602$, which was observed in binary particle packing~\cite{Ferraz2021} with almost the same particle population density and size ratio $\lambda=a/b$. A possible explanation for the notable difference compared to binary packs is the presence of structural constraints observed in particles with $\Sigma_{2v}(2\pi/3)$- arrangement, i.e., bent structure. These constraints introduce additional void spaces among the particles and promote interlocking, resulting in enhanced rolling resistance.
	
	Fig.~\ref{fig:05} displays the mean coordination number $z$ of the particles as a function of time. The mean coordination number $z$ is the number of neighboring particles that touch a given particle. A neighboring particle is considered when the bond distance between two primary spheres from different particles is equal to the sum of their radii, with an additional offset of $0.15\,\mu m$. Surprisingly, the mean coordination number is found to have an opposite behavior to the packing density, fairly increasing with increasing interaction strength $\varepsilon$. The final $z$ values obtained were as follows: 9.3(7) for the $\varepsilon=0$ case, 9.8(7) for the $\varepsilon=25\,\mu J$ case, and 10.3(2) for the $\varepsilon=50\,\mu J$ case. Despite the long-range interaction forces hindering particle packing, they clearly contribute to an improvement in the average compaction among the particles, resulting in more contact points among them.
	
	The packing quantities $\phi$ and $z$ are plotted as a function of the interaction strength $\varepsilon$ in Figs.~\ref{fig:06} and \ref{fig:07}, respectively. The red straight lines in these figures are linear regressions on the corresponding data. We obtained slopes of $4.10 \times 10^{-3}$ and $1.93 \times 10^{-2}$ for the quantities $\phi$ and $z$, respectively. Therefore, we can conclude from these results that packing quantities such as $\phi$ and $z$ are quite sensitive to long-range interaction forces, and do not exhibit shielding effects against such forces as found in binary particle packing with the same particle population density and size ratio $\lambda=a/b$.
	
	As a way to investigate the random packing structure formed, we have calculated the radial distribution function (RDF)~\cite{Rapaport1995} of the samples formed. It can be understood as the probability of finding one particle at a given distance from the center of a reference particle. The RDF is defined as
\begin{equation}\label{eq:23}
 g(r_{i})=\dfrac{n(r_{i})}{4\pi \, r_{i}^{2} \, \delta r_{i}\, \rho},
\end{equation}
where $n(r_{i})$ the number of particle centers within the $i$-th spherical shell of radius $r_{i}$ and thickness $\delta r_{i}$, and $\rho$ is the number of centers of primary spheres per volume. In the above equation, we set $\delta r_{i}=0.01 \, \mu m$, and regard a number of spherical shells $N_{r}=100$ for $g(r)$ computation. Fig.~\ref{fig:08} shows typical RDFs as a function of the radial distance for the random packing structures by considering two different $\varepsilon$ values, namely $\varepsilon=0\, \mu J$ (absence of long-range interactions) and $\varepsilon=50\, \mu J$. These curves were obtained by averaging the individual RDFs of all particles inside the bulk region of the formed particle aggregate. This bulk region is defined here as a smaller virtual box centered at the central point of the aggregate and having an offset distance of $1.0\;\mu m$ from each actual wall of the confinement box. From Fig.~\ref{fig:08}, one can observe four main peaks in RDF for $\Sigma_{2v}(2\pi/3)$-triplets packings. Such peaks are localized around the distances $0.20\, \mu m$, $0.23\, \mu m$, $0.30\, \mu m$ and $0.40\, \mu m$. These peaks correspond to different contact types between primary spheres. The main peak observed at $0.23, \mu m$ is attributed to the triplet structure itself. While the first, third, and fourth peaks correspond to contacts between primary spheres with radii: $a = 0.10 \, \mu m$; $a = 0.10 \, \mu m$ and $b = 0.20 \, \mu m$; and $b = 0.20 \, \mu m$, respectively. It can be seen that the increase in long-range forces, promoting compaction among the particles, has led to a significant rise in the RDF profile for the case with $\varepsilon=50\, \mu J$, compared to the scenario where long-range interactions are absent (case $\varepsilon=0\, \mu J$).

In addition, it is also important to understand how the long-range forces change the orientational order of the formed structures. To accomplish that, let us first define a local structural parameter sensitive to ordering as
\begin{equation}\label{eq:24}
\psi _{i,m}^l  = \frac{K}{{n_b }}\sum\limits_{j = 1}^{n_b} {Y_{lm} (\theta _{ij} ,\varphi _{ij} ),} 
\end{equation}
where $\psi _i ^l$ is a complex vector with $|m|\leq l$ components assigned to every particle $i$ in the system. In Eq.\eqref{eq:24}, the sum runs over all $n_b$ nearest neighbors of the particle $i$ and $K$ is a normalization constant so that the complex inner product $\sum\limits_m^{} {\psi _{i,m}^l} \,\psi _{i,m}^{l*}  = 1$. For a given pair of particles ($i$,$j$), $Y_{lm} (\theta _{ij} ,\varphi _{ij})$ represents the spherical harmonic associated with the bond vector $\vec r_{ij}$ connecting the centers of mass of this pair, where $\theta _{ij}$ and $\varphi _{ij}$ are the corresponding polar angles of this vector relative to a fixed coordination system. In order to check for both cubic symmetry (e.g., fcc and bcc structures) and icosahedral symmetry, we have taken $l=6$ (and $-6\leq m \leq 6$) in Eq.~\eqref{eq:24}. Thus we deal with $N$ 13-element complex vectors for bond-ordering analysis.

Once the order parameter $\psi _i ^l$ has been defined, further information about the bond-orientational order can be obtained from the orientation pair correlation function given by
\begin{equation}\label{eq:25}	
G _{6} (\vec r) = \sum\limits_{m} {<\psi _{i,m}^{l = 6} (\vec 0)\,\psi _{j,m}^{l = 6} (\vec r)^* >}.
\end{equation}
   
In the above equation, the angular brackets indicate an average over all particles separated by $\vec r$ in the bulk. It is known that the ``bond" between any two particles $i$ and $j$ is recognized as crystal-like if $G _{6} (\vec r)>0.5$~\cite{Steinhardt83,Desmond2009}. Thus we can check for crystallization, as well as changes in the ordering of the particles, due to the action of the long-range forces by computing $G _{6}$ as a function of the radial distance between particles. 

In Fig.\ref{fig:09} is shown the orientation pair correlation function $G_{6}$ as a function of the radial distance for random packing structures formed by $\Sigma_{2v}(2\pi/3)$-triplets, considering two different $\varepsilon$ value. It can be observed that the maximum bond orientation order ($G_6$) occurs around $0.45\, \mu m$, while the minimum $G_6$ value is around $0.60\, \mu m$. For both $\varepsilon$ cases, $G_{6}$ converges to an asymptotic average value of $0.086$. Besides, it is interesting to observe the overall effect of long-range forces on the bond-orientational order of the samples. For $\Sigma_{2v}(2\pi/3)$-triplet samples, the impact of these forces has, on average, reduced their bond-orientational order.

Another important question that may arise is how the various forces acting on each particle are distributed in the random packing structures that are formed. Because the contact and LJ forces differ from gravitational force in the sense that they have no preferred direction and are randomly oriented in such structures. Here we consider two force ratios: 
\begin{equation}\label{eq:26}
	{\zeta _{n}} = \frac{{\left| {\sum\limits_j {\vec F_{ij}^{n}} } \right|}}{{\left| {{m_i}\vec g} \right|}}
\end{equation}
and
\begin{equation}\label{eq:27}
	{\zeta _{LJ}} = \frac{{\left| {\sum\limits_j {\vec F_{ij}^{LJ}} } \right|}}{{\left| {{m_i}\vec g} \right|}},
\end{equation}
 where $\zeta _{n}$ and $\zeta _{LJ}$ represent the ratios of the magnitudes of the normal contact force and the LJ force, respectively, to the magnitude of the gravitational force. Figs.\ref{fig:10} and \ref{fig:11} show the force probability distributions in the random close-packing structures with $\Sigma_{2v}(2\pi/3)$-triplets for different $\varepsilon$ magnitudes. The normal contact force distribution is long-tailed and unimodal, spanning a wide range from $\zeta _{n}=0$ to over $\zeta _{n}=100$ for all $\varepsilon$ cases shown.  Whereas the LJ force distribution is bimodal-like and significantly narrower, ranging from $\zeta _{LJ}=0$ up to $\zeta _{LJ}=5$ for the case when $\varepsilon=25 \, \mu J$, and from $\zeta _{LJ}=0$ up to $\zeta _{LJ}=10$ for the case when $\varepsilon=50 \, \mu J$.

\section{\label{sec:c} Conclusions}

In this study, DEM simulations were performed to study the random close packing of $\Sigma_{2v}(2\pi/3)$-triplets at micrometer scales. Both contact forces and long-range dispersive forces were taken into account in these simulations. Several cases of random packing with triplets were treated by varying the long-range interaction strength ($\varepsilon$) during the construction of the samples. The packing dynamics was studied by evaluating over time different physical observables, including the packing density, mean coordination number, and average ratio of the kinetic energy to the gravitational potential energy. The radial distribution and orientation pair correlation functions were computed to characterize the particle structures formed over different values of the $\varepsilon$ control parameter. In addition, the force distributions in the random packing structures of $\Sigma_{2v}(2\pi/3)$-triplets were also analyzed. It was found that both the packing dynamics and the final values of the observables are highly sensitive to variations in the $\varepsilon$ values.

It was found that the packing density of the formed samples decreased with increasing interaction strength ($\varepsilon$). Surprisingly, the mean coordination number of the samples exhibited an opposite behavior to the packing density, increasing slightly with higher interaction strength. The possible explanation for such behavior is that even though the long-range interaction forces hinder particle packing, they clearly contribute to improving the average compaction among the particles, resulting in more contact points among them. This is supported by the RDF profiles and the contact force distribution of the samples, which showed a significant increase compared to the scenario without long-range interactions. Furthermore, it was also observed that the presence of long-range forces, on average, reduced the bond-orientational order of the formed samples. The distribution range of the normal contact force was found to be larger than that of the LJ force. The probability distribution of the normal contact force follows a long-tailed unimodal pattern, while that of the LJ force displays a bimodal form. 

Finally, it is important to stress that the present results obtained through particle sedimentation mechanism may be different from those obtained by using other methods. By changing the protocol for generating such particle aggregates, one may obtain slightly different results. The present study is important due to its potential applications in the development and manufacture of new materials, such as metallic alloys and ceramic compounds. Additionally, it contributes to the modeling of the atomic structure of amorphous metals composed of atoms of similar sizes.

\section{Acknowledgements}
We wish to thank UFERSA for computational support.

\section{\label{sec:ref} References}



\bibliographystyle{elsarticle-num-names} 

\end{document}